\documentclass[prd,twocolumn,aps,showpacs,nofootinbib,nobibnotes,superscriptaddress,preprintnumbers]{revtex4}
\usepackage{epsfig}
\usepackage{amssymb}
\usepackage{graphics}
\usepackage{bm}

\newcommand{\ff}{\bm{f}}
\newcommand{\xx}{\bm{x}}
\newcommand{\nab}{\mbox{\boldmath $\nabla$} {}}
\newcommand{\ppsi}{\mbox{\boldmath $\psi$} {}}
\def\urms{u_{\rm rms}}
\def\cs{c_{\rm s}}
\newcommand{\Fig}[1]{Fig.\ \ref{#1}}
\newcommand{\FFig}[1]{Figure \ref{#1}}
\newcommand{\Sec}[1]{Sec.\ \ref{#1}}
\def\EEK{{\cal E}_{\rm K}}
\def\EEM{{\cal E}_{\rm M}}

\def\Rey{\mbox{\rm Re}}
\newcommand{\yprd}[3]{, Phys.\ Rev.\ D {\bf #2}, #3 (#1).}
\newcommand{\ypre}[3]{, Phys.\ Rev.\ E {\bf #2}, #3 (#1).}
\newcommand{\yprl}[3]{, Phys.\ Rev.\ Lett.\ {\bf #2}, #3 (#1).}
\newcommand{\yjfm}[3]{, J.\ Fluid Mech.\ {\bf #2}, #3 (#1).}

\def\half{{1\over2}}

\begin{document}

\title{Evolution of inflation-generated magnetic field through phase transitions}

\date{\today} 
\preprint{NORDITA-2012-45, KSUPT-12/2}

\author{Tina Kahniashvili}
\email{tinatin@phys.ksu.edu}
\affiliation{The McWilliams Center for
Cosmology and Department of Physics, Carnegie Mellon University,
5000 Forbes Ave, Pittsburgh, PA 15213, USA}
\affiliation{Department of Physics, Laurentian University, Ramsey
Lake Road, Sudbury, ON P3E 2C, Canada} \affiliation{Abastumani Astrophysical Observatory, Ilia State
University, 3-5 Cholokashvili Ave, Tbilisi, GE-0194, Georgia}

\author{Axel Brandenburg}
\email{brandenb@nordita.org}
\affiliation{Nordita, KTH Royal Institute of Technology and Stockholm University,
Roslagstullsbacken 23, 10691 Stockholm, Sweden}
\affiliation{Department of Astronomy, AlbaNova University Center,
Stockholm University, 10691 Stockholm, Sweden}

\author{Leonardo Campanelli}
\email{Leonardo.Campanelli@ba.infn.it}
\affiliation{Dipartimento di Fisica, Universit\`{a} di Bari, I-70126 Bari, Italy}

\author{Bharat Ratra}
\email{ratra@phys.ksu.edu}
\affiliation{Department of Physics,
Kansas State University, 116 Cardwell Hall, Manhattan, KS 66506, USA}

\author{Alexander G.\ Tevzadze}
\email{aleko@tevza.org}
\affiliation{Faculty of Exact and Natural
Sciences, Tbilisi State University, 1 Chavchavadze Ave., Tbilisi,
0128, Georgia}

\begin{abstract}
We study the evolution of an inflation-generated
magnetic field, due to its coupling to fluid motions, during
cosmological phase transitions. We find that the magnetic field stays
almost unchanged on large scales, while on small scales the spectrum
is modified in such a way that power at small scales becomes
progressively suppressed. We also show that the magnetic field generates
turbulent motions in the initially turbulence-free plasma.
On large scales, the slope of the resulting kinetic energy
spectrum is consistent with that of white noise.
\end{abstract}


 \maketitle

\section{Introduction}

The origin of the coherent large-scale ($\sim$ 10\,kpc) part of
galactic magnetic fields, of $\mu$G strength, is under active discussion
\cite{Subramanian:2010ab, Kandus:2011ab, Widrow:2012ab, Yamazaki:2012ab}.
On larger, Mpc scales, until recently there were only upper limits,
the most restrictive being of order a few nG depending on the observational
technique used to measure the intergalactic magnetic field strength
\cite{Kahniashvili:2010ab}. Recently, there have been a
number of published lower limits on a putative large-scale magnetic
field of strength $10^{-1 \pm 1}$\,fG (1\,fG = $10^{-15}$\,G)
\citep{Neronov:2010ab}, or possibly two orders of magnitude smaller
\citep{Dermer:2011ab}.\footnote{
These techniques for limiting a large scale cosmological magnetic
field might be unreliable \cite{Broderick:2011ab}, but see their
Sec.\ 4 where they note that more work will be needed to firm up
these arguments and to determine whether the techniques
used to establish the lower limits are indeed unreliable.}

Almost certainly, the galactic fields are the amplified remnants of
significantly weaker ``seed'' magnetic fields. Quantum mechanical
fluctuations during inflation \cite{Fischler:1985ab} is a leading
candidate for generating the needed seed magnetic field
\cite{Turner:1988ab, Ratra:1992ab, Ratra:1991ab}.
To generate a large enough seed magnetic field through quantum mechanical
fluctuations during inflation, conformal invariance must be broken during
inflation. A simple, realistic, illustrative model couples the
abelian vector field with field strength tensor $F_{\mu\nu}$ to
the scalar inflaton field $\phi$ through a dilaton-like coupling,
generalizing the Maxwell lagrangian density to $e^{\alpha\phi} F_{\mu\nu}
F^{\mu\nu}$ where $\alpha$ is a parameter \cite{Ratra:1992ab, Ratra:1991ab}.
In the case of power-law inflation, and depending on the value of $\alpha$,
this can result in a large enough seed magnetic field to explain the
observed galactic magnetic fields. This is an observationally viable model.
For a more detailed description of the model see Sec.\ II below.

After the end of inflation, such an inflation-generated magnetic field
will be correlated over super-Hubble-radius scales. It would induce
observable signatures in the cosmic microwave background (CMB)
radiation anisotropies at the epoch of recombination (the last scattering
surface) if its current amplitude on Mpc scales is of the order of a nG
\cite{CMBanisotropy}.\footnote{
The effects of an homogeneous magnetic field on the CMB anisotropy,
and the resulting non-Gaussianity, are discussed in Refs.\
\cite{CMBnongaussianity}.}
The properties of an inflation-generated primordial seed magnetic
field depend on the parameters of the inflation model. If cosmological
observations confirm the presence of an inflation-generated
magnetic field, these measurements could be used to probe the physical
conditions during inflation, including the shape of the inflaton
potential energy density as well as the coupling between the inflaton
and the vector gauge field.

To check the consistency of the model, the  primordial magnetic field shape and
amplitude should be measured in as many ways as possible. The simplest limit
arises from the cosmological expansion dynamics during big bang
nucleosynthesis. This requires that the energy density of the magnetic field
should not be larger than about 10\% of the radiation energy density. This limits
the present (inflation-generated) magnetic field strength to less than a
few $\mu$G, if the  primordial magnetic field was generated prior to or
during big bang nucleosynthesis, and was not damped or amplified by a
magnetohydrodynamic (MHD) or some other process and so stays frozen into the
plasma \cite{Kawasaki:2012ab}.

In addition to the CMB temperature anisotropies that a primordial magnetic
field induces (as mentioned above), such a field will Faraday-rotate
the CMB polarization anisotropies \cite{Faraday}. Currently available
Faraday rotation data give a bound on the primordial magnetic field strength
of less than a few nG (for a scale-invariant or homogeneous
primordial magnetic field).

Another interesting signature of a cosmological magnetic field is the
relic gravitational wave signal generated by the anisotropic magnetic
stress \cite{Deriagin:1987ab}. The amplitude of the induced gravitational
waves is determined by the magnetic field energy density, so a direct
measurement of the resulting gravitational wave signal can lead to an
independent limit on the magnetic field strength; see Ref.\
\cite{Wang:2010ab} and references therein.

After the Universe reheats at the end of inflation, the plasma
that was created then has large conductivity and it is conventional to
assume that this remains the case as the Universe evolves to the
present. In this case the large-scale cosmological magnetic field
behaves as a frozen-in field with an evolution determined by the
simple, flux-conservation, dilution of magnetic field lines,
${\bf B}({\bf x}, t) \propto {\bf B_0}({\bf x})/ a^2(t)$, where
$t$ is the physical cosmic time and $a(t)$ is the cosmological scale
factor. On the other hand, the evolution of a primordial magnetic
field is a complex process influenced by MHD as well as by the
dynamics of the Universe \cite{axel1,banerjee,campanelli2004,
campanelli2007}. In particular, the presence of a magnetic field
can dramatically affect primordial turbulence (e.g., when the
turbulence is associated with cosmological phase transition bubble motions)
\cite{axel1,axel-2,caprini09}. Furthermore, the presence of a magnetic field
itself might lead to the development of turbulent motions, and so
affect the turbulence \cite{B,axel}.

In a recent examination \cite{kbtr10} of the effects of the MHD
coupling between a primordial magnetic field and turbulence during
a cosmological phase transition, we considered two different initial
shapes for the spectrum of the primordial magnetic field: a single-scale
magnetic field and a magnetic field
with a Batchelor spectrum at large scales. In this paper we present a
similar analysis for modified initial conditions for an inflation-generated
primordial magnetic field \cite{Ratra:1992ab}, coupled via the usual
MHD equations with the fluid, during the electroweak or QCD phase
transitions. We consider both non-helical and helical magnetic field cases.

We assume that the phase transition bubbles induce a typical length scale
at which the magnetic field starts to interact with the phase transition fluid.
The relevant difference between the electroweak and QCD phase transitions
is encoded in the difference between values of parameters such as the
temperature $T_\star$, the number of relativistic degrees of freedom
$g_\star$, the bubble number, and bubble sizes. We assume initial absence
of primordial turbulence, i.e., we assume that the plasma is initially at
rest (although it is possible to generate turbulent motions through bubble
collisions and nucleation \cite{kos}). The characteristic parameter of
the primordial magnetic field is the r.m.s.\ Alfv\'en velocity
$v_A = {B}/{\sqrt{16\pi {\rho_{\rm rad}}/3}}$. Here,
$\rho_{\rm rad} \simeq \rho_{\rm thermal}$ is the radiation energy
density.
We use $\Omega_{\rm rad} h_0^2= 2.56 \times 10^{-5} $, where
$\Omega_{\rm rad}$ is the radiation energy density parameter and $h_0$
is the Hubble constant in units of 100 km s$^{-1}$ Mpc$^{-1}$, for
a current CMB temperature $T_0=2.74\,$K.
At temperature $T_\star$,
$\rho_{\rm rad}(T_\star) = {\pi^2}g_* (T_*)^4/30 $.
The Alfv\'en velocity does not depend on $T_\star$ but is weakly
dependent on $g_\star$, i.e., $v_A \propto g_\star^{-1/6}$.

In our previous simulations \cite{kbtr10} we studied
phase-transition-generated magnetic fields coupled to a
relativistic fluid and discovered that equipartition between kinetic
and magnetic energy densities is reached within reasonably short times.
In this paper our main purpose is to consider different initial conditions.
In particular, in the case when the magnetic field is generated during
inflation, we investigate the kinds of turbulent motions that result
from the coupling of the magnetic field with the fluid, and determine
how this affects the evolution of the field itself. We show that the
presence of a magnetic field on large scales ensures a rapid rise of
the velocity field on large scales. On the other hand, magnetic field
decay on large scales occurs at slow rates.

The structure of our paper is as follows. In Sec.\ II we briefly
describe magnetic field generation during inflation. In Sec.\
III we discuss the phenomenological coupling of the magnetic field
to the turbulent plasma. In Sec.\ IV we present our numerical
simulation results. We discuss our results and conclude in Sec.\ V.

\section{Inflation-generated magnetic field}

\subsection{Non-helical magnetic field}

Any field included in the lagrangian density during inflation will be produced
by quantum-mechanical fluctuations. These fluctuations are then stretched
by the cosmological expansion, leaving the Hubble radius during inflation
and re-entering it at much later times.

Adding the standard abelian vector field lagrangian density  $F_{\mu\nu}
F^{\mu\nu}/{\hat e}^2$ (where $\hat e$ is the vector potential coupling
constant) to the usual general relativity and scalar inflaton field
lagrangian density results in electric and magnetic fields being
generated during inflation.
Of course, these are not the usual low-energy electric and magnetic
fields; rather, at the very least, they are hypercharge electric
and magnetic fields.
During inflation the conductivity vanishes, but particle creation at
reheating quickly turns the Universe into a very good conductor at the
end of inflation. This short circuits the electric field but does not affect
the large-scale magnetic field. For a close to de Sitter exponentially
expanding epoch of inflation, on a scale of a thousandth of a Hubble
radius now (i.e., a few Mpc now), the r.m.s.\ magnetic field strength
now is of order $10^{-59}$ G and very insignificant \cite{Ratra:1992ab}.
In this case the power spectrum of the magnetic field is
$|B(k)|^2 \propto k/a^4$,
where $k$ is the coordinate wave number, and the evolution with scale factor
$a$ is as expected from flux conservation.
This tiny value is a consequence of the conformal invariance of the
abelian vector field lagrangian density and the background spacetime.
To generate a large enough magnetic field requires using a model
in which conformal invariance is broken \cite{Turner:1988ab, Ratra:1992ab,
Ratra:1991ab}.

A simple way to break conformal invariance is to couple the inflaton
scalar field $\phi$ to the abelian vector field through a dilaton-like
coupling, generalizing the Maxwell lagrangian density to
$e^{\alpha\phi} F_{\mu\nu} F^{\mu\nu}/{\hat e}^2$ ($\alpha$ is a
parameter and the exponential form of the dilation-like coupling
was chosen for simplicity) \cite{Ratra:1992ab, Ratra:1991ab}.
At the end of inflation, during reheating, $\phi$ freezes at its
vacuum expectation value and the usual abelian vector field lagrangian
density is recovered. During inflation however, the model behaves
as though it has a varying abelian vector field coupling constant.
This modifies the scale factor and wave number dependence of the
generated magnetic field. For power-law inflation \cite{Ratra:1989ab},
power is shifted to the infrared and the magnetic field energy
density redshifts slower than the usual uncoupled abelian vector
field case.

Depending on the value of $\alpha$, this can result in a large enough
seed magnetic field to explain the observed galactic magnetic fields.
The case of greatest interest is near the limit of de Sitter exponential
expansion inflation, which results in a close to scale-invariant
Harrison-Peebles-Yu-Zeldovich spectrum of energy density perturbations
(consistent with the observational indications, see, e.g., Ref.\
\cite{Ratra:2008ab}) with a current epoch magnetic field scale-invariant
power spectrum $|B(k)|^2 \propto k^{-3}/a^4$, or
$\langle B^2(r)\rangle^{1/2} \propto r$,
and r.m.s.\ amplitude of order a few nG on a scale of a thousandth of a Hubble
radius \cite{Ratra:1992ab, Ratra:1991ab}. These computations have been
carefully checked, confirmed, and extended \cite{Bamba:2004ab}. This is
an observationally viable model.

In this model, there are regions in model parameter space where the vector
field fluctuations during inflation are large enough to invalidate the linear
perturbation assumption, but this does not occur in the region of parameter
space where a strong-enough current-epoch large-scale magnetic field is
generated \cite{Ratra:1992ab, Ratra:1991ab, Demozzi:2009ab, Kanno:2009ab}.
That is, backreaction is not a significant issue for this classically
consistent model. While the abelian vector field
coupling becomes large during inflation, this is not important for the
phenomenological, effective classical model \cite{Ratra:1992ab,
Ratra:1991ab, Demozzi:2009ab, Subramanian:2010ab, Caldwell:2011ab}.
Instead, much like the case of the ``standard'' $\Lambda$CDM cosmological
model, it is of great interest to try to find a more fundamental,
quantum-mechanically consistent, model that can give rise to this
classical, observationally-successful, inflationary magnetogenesis model.

An extension of this model, with two scalar fields instead of one,
is also viable \cite{Lemoine:1995ab}. Here, in addition to the usual
scalar inflaton field, a new scalar field, coupled as a dilaton to
the abelian vector field, is introduced. Alternatively, as discussed
next, the second field is taken to be a pseudoscalar, coupling like
an axion to the abelian vector field. For other inflationary
magnetogenesis variants see Refs.\ \cite{rest}.

\subsection{Helical magnetic field}

The inverse cascade scenario, a much discussed magnetic field amplification
mechanism, requires non-zero magnetic helicity, if it is to
be able to transfer magnetic power from small scales to large
scales and so amplify the large-scale magnetic field
\cite{Brandenburg:2012ab}. It is therefore of interest to consider
inflationary magnetogenesis models that can also generate
magnetic helicity.

In a Friedmann-Lema\^{\i}tre-Robertson-Walker cosmological model,
magnetic helicity is \cite{Campanelli2}
\begin{equation}
\label{H1} H_B(t) = \int \! d^3 x \, {\textbf A} \cdot \nabla
\times {\textbf A},
\end{equation}
where the vector potential ${\textbf A}$ is related to the magnetic
field through $a^2 {\textbf B} = \nabla \times {\textbf A}$. Magnetic
helicity is usually associated with a nontrivial configuration of
the magnetic field where the magnetic flux tubes are twisted and/or
linked \cite{B}. Magnetic helicity resembles a Chern-Simons term,
for its time variation is given by $H_B(t_2) - H_B(t_1) =
\frac{1}{4} \! \int_{t_1}^{t_2} \! d^4 x F_{\mu \nu}
\widetilde{F}^{\mu \nu}$~\cite{Campanelli2}, where
$\widetilde{F}^{\mu \nu}$ is the vector field strength tensor dual.
This similarity is strengthened by the fact that a helical magnetic field
may be identified as the projection of a non-abelian gauge field
configuration carrying a nonvanishing Chern-Simons number \cite{Jackiw},
and that the magnetic helicity coincides with this winding number.

Moreover, helical magnetic fields possess a number asymmetry
between left-handed and right-handed abelian vector field helicity
states \cite{Campanelli2}. To see this it is useful to go to Fourier
space and introduce the orthonormal helicity basis
$\{{\textbf e}_{+}, {\textbf e}_{-}, {\textbf e}_{3} \}$,
with ${\textbf e}_{\pm} = ({\textbf e}_{1} \pm i {\textbf e}_{2})/\sqrt{2}$
and ${\textbf e}_{3} = {\textbf k}/k$,
such that $\{{\textbf e}_{1}, {\textbf e}_{2}, {\textbf e}_{3} \}$ form
a right-handed orthonormal basis. Decomposing the magnetic field in
polarization states as ${\textbf B}({\textbf k}) = B_{+} {\textbf e}_{+}
+ B_{-} {\textbf e}_{-}$, the magnetic helicity reads
\begin{equation}
\label{H2}
H_B(t) = \frac{a^4}{2\pi^2} \int \! dk \, k \left(|B_{+}|^2 -
|B_{-}|^2 \right).
\end{equation}
It is clear, accordingly, that a helical magnetic field possesses a
nonzero difference between the number of left- and right-handed
abelian vector potential polarization states.

After reheating, the Universe is a good conductor and so if a helical
magnetic field is created during inflation its magnetic helicity survives
to the present. Since magnetic helicity is odd under discrete $P$ and
$CP$ transformations, a cosmological helical magnetic field would be a
signature of macroscopic $P$ and $CP$ violation.

Models for generating helical magnetic fields in the early Universe
exist in the literature~\cite{Turner:1988ab,Garretson,Field,Prokopec,
Campanelli3,Jackiw,Helical}. In order to generate a helical magnetic
field during inflation, one can add an interaction term
of the form $I(\phi) F_{\mu\nu} \widetilde{F}^{\mu\nu}$ to the abelian
vector potential lagrangian density. Here $I(\phi)$ is a pseudoscalar
function of some dynamical or background field $\phi$.

A coupling $I(\phi) \propto \phi$, between the abelian vector potential and
the axion field $\phi$ was studied in Refs.\ \cite{Turner:1988ab,Garretson,
Field}, while in Ref.\ \cite{Prokopec} the pseudoscalar $\phi$ was assumed
to drive inflation. In these cases, because of the extra derivative compared
to the $e^{\alpha\phi} F_{\mu\nu} F^{\mu\nu}/{\hat e}^2$ term, magnetic
power is concentrated on smaller scales, with insignificant magnetic
power on cosmological scales of interest.

In Ref.\ \cite{Campanelli3}, to produce larger-scale magnetic field power,
the function $I$ was taken to be a time-dependent function peaked at long
wavelengths (this particular coupling to the abelian vector potential
could be realized by a tachyonic massive pseudoscalar field or a massless
pseudoscalar field non-minimally coupled to gravity). It was shown that,
depending on the strength of the coupling, a maximally helical field with
a scale-invariant spectrum, $|B(k)|^2 \propto k^{-3}$, could be produced
as an excitation of the vacuum during inflation. Although it can be quite
large, large enough to act as a seed for the magnetic fields we observe
today in galaxies and galaxy clusters, as before, its backreaction on
the development of inflation is completely negligible.

\section{Coupling of Magnetic Field and Turbulent Motions}

Because of conformal invariance, the usual flat
spacetime relativistic MHD equations are identical to the MHD
equations in an expanding Universe with zero spatial curvature
when physical quantities
are replaced by their co-moving counterparts and conformal time
${\eta}$ is used in place of physical time \cite{enqvist}. Based on this fact,
we perform direct numerical simulations of MHD turbulence in an
expanding Universe using the usual flat spacetime MHD equations
with a relativistic equation of state.
Note that our simulations are based on the relativistic
equations even when studying evolution of turbulence with
non-relativistic bulk velocities.

Let us briefly describe the primordial magnetic field coupling to the
fluid. As noted above, the typical characteristic length scale of this
coupling is the phase transition bubble size, $\lambda_0 = \gamma
\lambda_H$, where $\gamma<1$ is a parameter connected with the number
of bubbles, $N$, within a Hubble radius, i.e., $\gamma^{-1} \propto
N^3$, and the Hubble radius
\begin{equation}
\lambda_{H} = 5.8 \times 10^{-10}~{\rm Mpc}\left(\frac{100\,{\rm
GeV}}{T_\star}\right) \left(\frac{100}{g_\star}\right)^{{1}/{6}}.
\label{lambda-max}
\end{equation}

As we noted above we will consider both electroweak and QCD phase
transitions. The phase transitions are characterized by different
maximal correlation lengths due to the difference in the $\gamma$
parameter (that is equal to 0.01 and 0.15 for electroweak and QCD phase
transitions, respectively) and in the Hubble radius; see
Eq.~(\ref{lambda-max}) (that is equal to 0.006\,$\mu$pc and 5.5\,pc for
electroweak and QCD phase transitions, respectively). As we noted
above, the initial value of the Alfv\'en velocity depends weakly
on $g_\star$ ($v_A \propto g_\star^{-1/6}$), which also makes
a difference between the electroweak and QCD phase transitions
of the order of unity (1.37), which we discard. The conductivity of the
Universe is high enough during the phase transitions: of course the
physical characteristics of the plasma (such as viscosity and
conductivity) depend on temperature and vary from 100\,GeV
(electroweak phase transition) to 0.15\,eV (QCD phase transition).
On the other hand, in our simulations we assume that, for the goal
of our study, the main difference between the electroweak and
QCD phase transitions consists only in a difference of the initial
correlation length. Our assumption can be justified as follows:
(i) the forcing amplitude at the initial moment weakly
depends on the relativistic degrees of freedom, $\propto
g_\star^{-1/6}$; (ii) the physical conditions of the Universe are
different, but new simulations for a wide range of the Prandtl numbers
show that the growth of the correlation length does not depend on the
value of the Reynolds number, as long as it is large enough \cite{future}.

We consider two types of forcings, irrotational and vortical,
and for both we assume that the forcing scale coincides with
the bubble size, $\lambda_0$. The corresponding wave number
is $k_0=2\pi/\lambda_0$.
Energy is being dissipated both viscously (characterized by the viscosity
$\nu$) and ohmically (characterized by the magnetic diffusivity $\lambda$,
which is inversely proportional to the conductivity).
Throughout this work we assume that the magnetic Prandtl number is
$\nu/\lambda=1$.
The smaller the value of $\nu$, the more extended the turbulent cascade,
and the larger the minimal mesh resolution required.
For all runs presented here we use $512^3$ mesh points.
The largest value of the Reynolds number based on the wavenumber $k_0$,
$\Rey\equiv\urms/\nu k_0$, is around 200 in all cases (here $\urms$
is the r.m.s.\ velocity).

There are many MHD simulations that use
vortical forcing \cite{axel-2} but fewer that use irrotational
forcing \cite{MB06}. However, we are unaware of any simulation
with an initial scale-invariant magnetic field spectrum,
$|B(k)|^2 \propto k^{-3}$, of appreciable strength.\footnote{
We require that the magnetic field satisfies the upper
limits from the current CMB and large scale structure observations,
\cite{CMBanisotropy} (that automatically satisfy the BBN limits),
so to be order of few nG. Such a magnetic field can be generated
during inflation \cite{Ratra:1992ab}. For helical magnetic
fields we assume that they are of maximal helicity
\cite{Campanelli3}.} We generate the corresponding magnetic vector
potential for such a field in Fourier space using modes with random
phases and suitable amplitudes, as was done in Ref.\ \cite{YHB04}
for the initial velocity field using the {\sc Pencil Code}
\cite{PC}, which is also used here.
We present magnetic and kinetic energy spectra normalized such that
$\int E_{\rm M}(k)\,dk=\langle{\textbf B}^2\rangle/2$ and
$\int E_{\rm K}(k)\,dk=\langle{\textbf u}^2\rangle/2$, respectively.

In all cases we keep the forcing amplitude at a fixed
level such that the resulting r.m.s.\ velocity is around $0.1\cs$,
where the constant $\cs$ is the isothermal sound speed.
The rms value of initial magnetic field strength, $B_{\rm rms}^{(0)}$,
is chosen such that $B_{\rm rms}^{(0)}/\cs$ is between 0.05 and 0.09.

In the following we describe the types of forcing applied to the system.

\subsection{Irrotational (potential) forcing}

We consider two types of irrotational forcings.
In both cases the forcing function is written as $\ff(\xx,t)=\nab\phi$,
where $\phi(\xx,t)$ is a random scalar function.
In the first case we model irrotational forcing in the form of spherical
expansion waves that reflect the dynamics of phase transition bubbles.
The forcing and some results for cases with no or weak magnetic fields
are described in Refs.\ \cite{DSB11,MB06}. The nondimensional radius of
the expansion waves is $k_1 R=0.133$, corresponding to a nominal forcing
wave number $k_0/k_1=15$. Here, $k_1=2\pi/L$ is the minimal wave number in
our computational cube of size $L^3$.
In the second case we use random plane waves with a wave number $k_0/k_1=30$.


\subsection{Vortical forcing}

Vortical forcing is accomplished by generating a random vector potential
$\ppsi$ such that $\ff=\nab\times\ppsi$.
Normally, $\ppsi$ is non-helical, but in some cases we have arranged
$\ppsi$ such that it consists of positively polarized waves \cite{axel-2}.
Like in the second case with potential forcing, the forcing wavenumber
is here chosen to be $k_0/k_1=30$.

\section{Results}

\subsection{Blob-like forcing}
\label{BlobLike}

Blob-like irrotational \cite{MB06} and plane wave \cite{Feder} forcings
have been used in the past to investigate
the production of vorticity through the viscous force,
or through the interaction of irrotationally-forced flow with global rotation
or shear with an isothermal equation of state, or through the baroclinic
term with the more general perfect gas equation of state \cite{DSB11}.
In the present case, again using an isothermal gas, the Lorentz force
associated with the initial magnetic field also produces vorticity \cite{Doso}.

\begin{figure}[t!]\begin{center}
\includegraphics[width=\columnwidth]{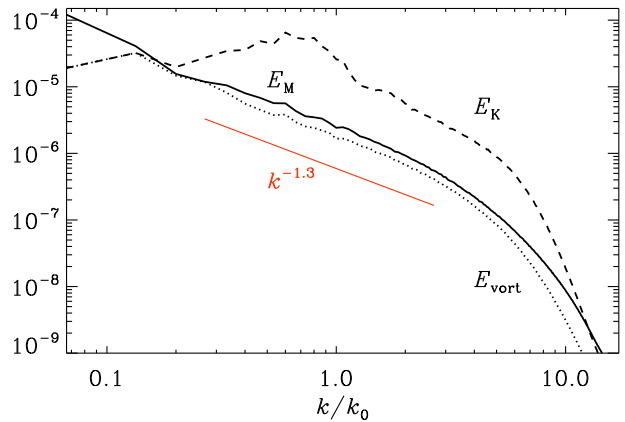}
\end{center}\caption[]{
Spectra of magnetic and kinetic energy as well as the
kinetic energy of the vortical component at the end of the run
for vortical blob-like forcing.
$\Rey\approx250$ and $B_{\rm rms}^{(0)}/\cs0.05$.
}\label{pkt512f_vort}\end{figure}

\begin{figure}[t!]\begin{center}
\includegraphics[width=\columnwidth]{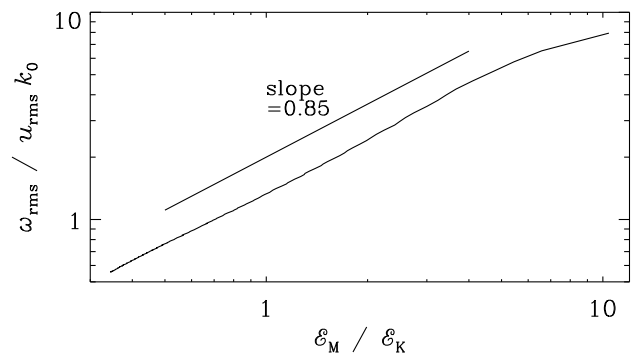}
\end{center}\caption[]{
Dependence of normalized r.m.s.\ vorticity on normalized magnetic energy density
during the decay for the run shown in \Fig{pkt512f_vort} for $\Rey\approx250$.
}\label{pvort512f}\end{figure}

In all cases investigated here, the magnetic energy density decays, so there is
either no dynamo action, or the initial magnetic field is still too
strong for dynamo action to occur because of excessive backreaction
on the flow via the Lorentz force. As the magnetic field decays, the
level of vorticity also decreases. In fact, at the end of the simulation,
the spectrum of the vortical part of the kinetic energy follows
closely that of the magnetic energy; see \Fig{pkt512f_vort}.
The r.m.s.\ vorticity turns out to be approximately proportional to the
magnetic energy density. Quantitatively, we find
$\omega_{\rm rms}/u_{\rm rms}k_0\approx0.1(\EEM/\EEK)^{0.85}$;
see \Fig{pvort512f}.
Here $ \omega_{\rm rms}$ is the r.m.s.\ vorticity and $\EEM$ and $\EEK$
are the magnetic and kinetic energy densities.
(We note that the exponent 0.85 is probably not robust; in another
simulation we have found a somewhat larger exponent $\approx1.2$,
for example.)

During the rising phase of the kinetic energy, the kinetic energy
spectrum is approximately proportional to $k^2$. At small scales,
we have an approximate $k^{-2}$ spectrum, in agreement with earlier
studies \cite{MB06,DSB11}. Here the spectral magnetic energy exceeds the total
spectral kinetic energy, and the vortical part of the kinetic energy spectrum
is only slightly below that of the total kinetic energy spectrum.

\subsection{Monochromatic vortical forcing}
\label{Monochromatic}

\begin{figure}[t]
\includegraphics[width=\columnwidth]{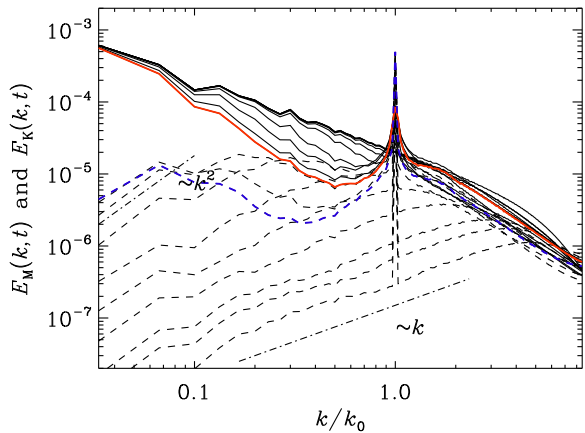}
\caption{
Magnetic (solid lines) and kinetic (dashed lines) energy spectra in
regular time intervals. $\Rey\approx170$.
The magnetic and kinetic spectra at the last time are additionally
marked in red and blue, respectively.
$\Rey\approx180$ and $B_{\rm rms}^{(0)}/\cs0.09$.
}\label{pkt1_u512_km1_b}
\end{figure}

Next we consider the case of monochromatic vortical forcing.
In the simulations presented here, the spectrum of kinetic energy is below
that of the magnetic energy, but it still shows a $k^2$ behavior
at small wave numbers and intermediate times.
At earlier times the spectrum is closer to a linearly increasing one;
see \Fig{pkt1_u512_km1_b}.
At intermediate times there is a characteristic decline of magnetic
energy at intermediate wave numbers ($k/k_0\approx0.3$), which then
leads to a similar decline of the kinetic energy at these wave numbers.

\begin{figure}[t]
\includegraphics[width=\columnwidth]{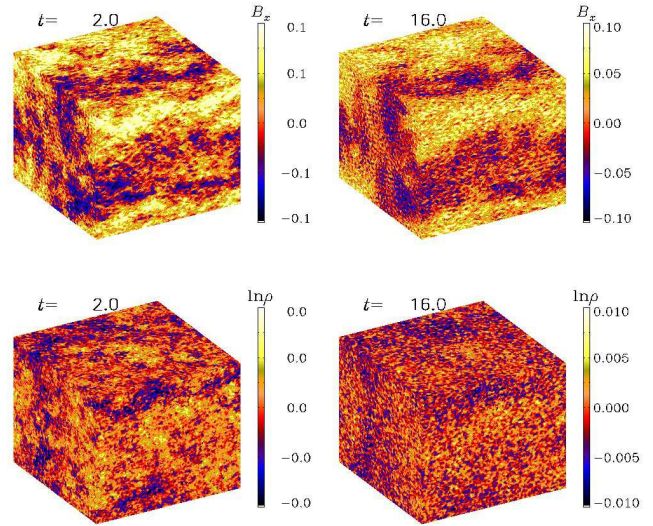}
\caption{
Visualization of the $B_x$ component and $\ln\rho$ for the run shown in
\Fig{pkt1_u512_km1_b}.
Note that at late times the small-scale velocity variations cause a
corresponding imprint on the magnetic field structure.
$\urms/\cs=0.05$ and $\Rey\approx180$.
}\label{u512_km1_b2}
\end{figure}

Visualizations of the $B_x$ component of the magnetic field and of the
logarithmic density, $\ln\rho$, are given in \Fig{u512_km1_b2}
for early and late times.
The initial $k^{-1}$ magnetic energy spectrum manifests itself in the
form of large random patches which also give an imprint on $\ln\rho$.
However, as the forcing proceeds, small-scale structures of scale
$\lambda_0$ become visible in $\ln\rho$ as well as in $B_x$.
This is quite different in the case of irrotational forcing,
of which we describe the results from a plane-wave forcing
formulation next.

\subsection{Irrotational plane-wave forcing}

\begin{figure}[t]
\includegraphics[width=\columnwidth]{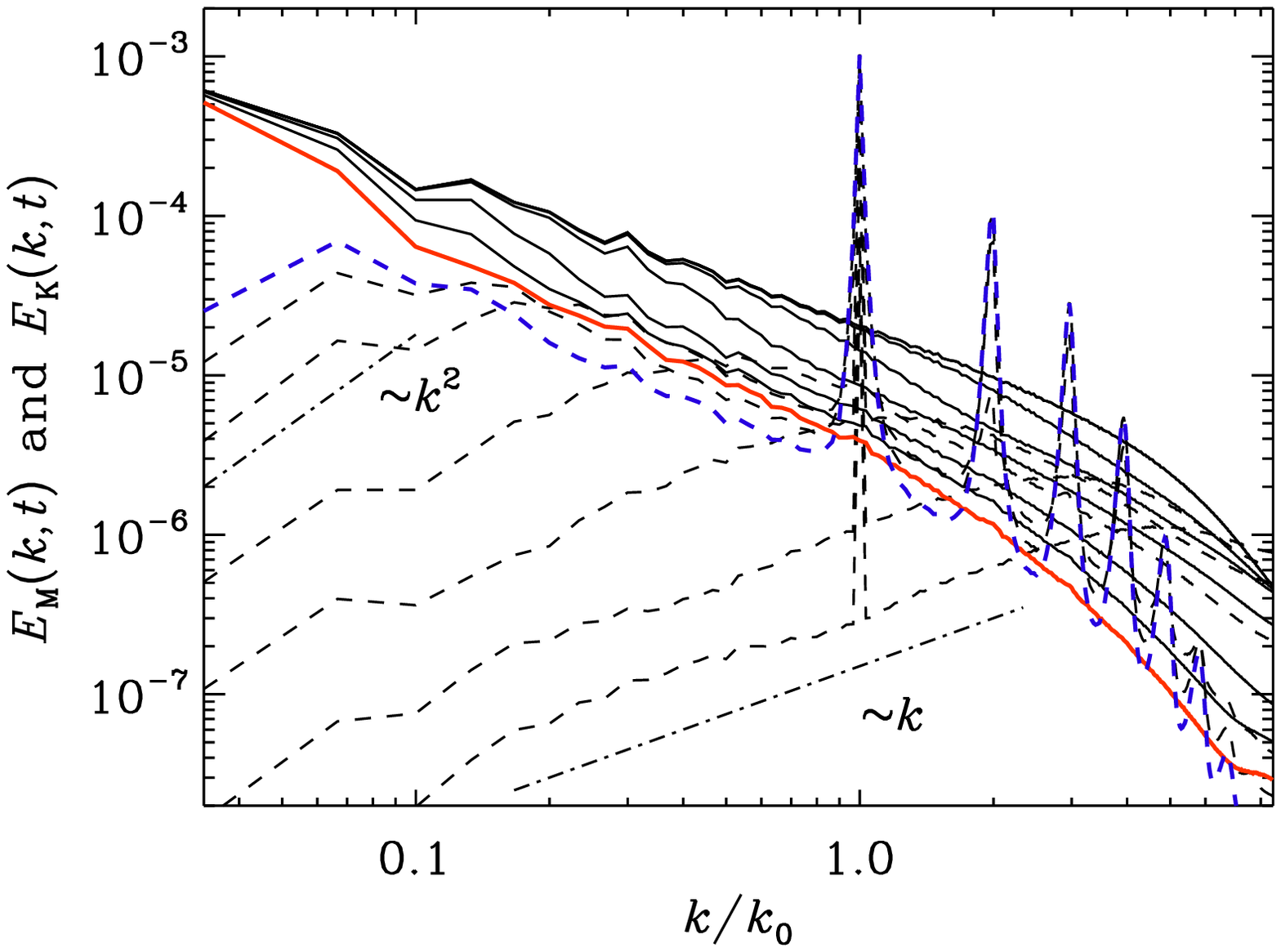}
\caption{
Similar to \Fig{pkt1_u512_km1_b}, but for potential forcing with plane waves.
$\Rey\approx220$ and $B_{\rm rms}^{(0)}/\cs0.09$.
}\label{pkt1_u512_km1_d}
\end{figure}

\begin{figure}[t]
\includegraphics[width=\columnwidth]{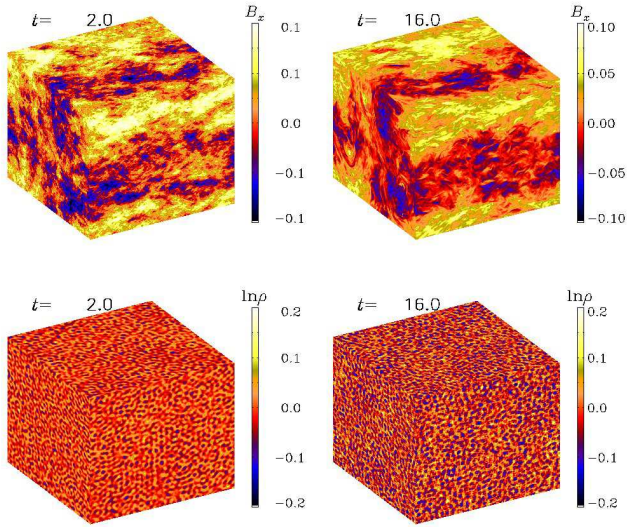}
\caption{
Visualization of the $B_x$ component and $\ln\rho$ for the run shown in
\Fig{pkt1_u512_km1_d}. Potential forcing (plane waves). Note that at late
times the small-scale velocity field has hardly any impact on the magnetic
field.
$\urms/\cs=0.07$ and $\Rey\approx220$.
}\label{u512_km1_d}
\end{figure}

The irrotational forcing considered in \Sec{BlobLike} is more realistic
for applications to phase transition bubbles than is plane wave forcing,
but to ease the comparison with the vortical forcing considered in
\Sec{Monochromatic} we now consider the case of plane-wave irrotational
forcing. In this case the spectra show initially the same behavior as
in the vortical case.
Even at late times there are similarities, except that the magnetic and
kinetic energy spectra now lack the characteristic decline that was visible
in the vortical case at $k/k_0\approx0.3$; see \Fig{pkt1_u512_km1_d}.
This is because the irrotational part of the flow does not interact
directly and sufficiently strongly with the magnetic field.
This is also clear from visualizations shown in \Fig{u512_km1_d} which
demonstrate that the small-scale structures of scale $\lambda_0$
leave no imprint on the magnetic field.

\begin{figure}[t]
\includegraphics[width=\columnwidth]{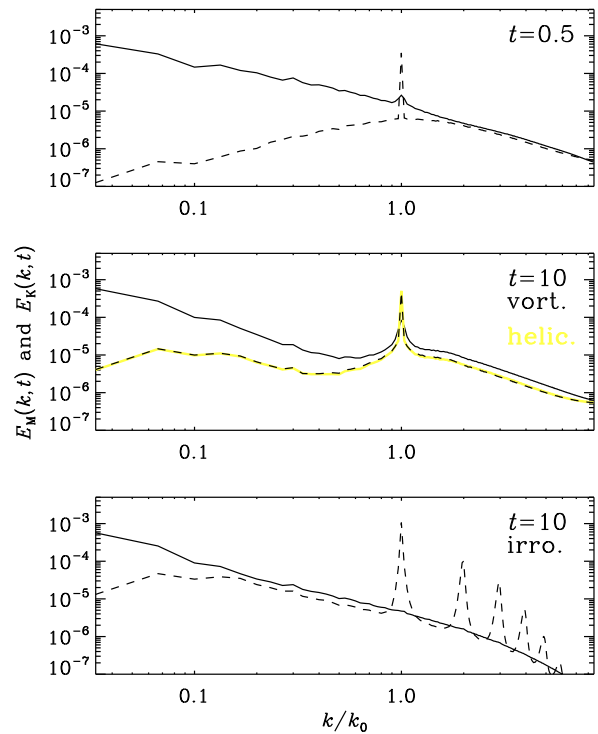}
\caption{
Comparison of magnetic (solid lines) and kinetic (dashed lines) energy
spectra at early and late times for vortical and irrotational plane-wave
forcing. The vortical case with helicity is shown in light yellow,
and nearly coincides with the non-helical case (dashed).
At early times the spectra are independent of the nature of the forcing.
}\label{pspec2tb}
\end{figure}

In \Fig{pspec2tb} we compare spectra of magnetic and kinetic energy at
early and late times for vortical and irrotational plane wave forcing. At
the early time the spectra are independent of the nature of the forcing, so
we show here, in the upper panel, only the vortical case at $t=0.5/\cs
k_1$. At a later time ($t=10/\cs k_1$), the kinetic and magnetic energy
spectra show a decline at intermediate wave numbers for the vortical case (see
middle panel), as already discussed above.
The result for the helical case is virtually indistinguishable.
This is because magnetic energy tends to drive the transfer of magnetic
energy to larger scales, but the spectral magnetic energy at those scales is
now already rather strong. In the irrotational case (see
bottom panel) the decline at $k/k_0\approx0.3$ does not exist.
This is because here vorticity is small and only this vortical
part of the flow interacts with the magnetic field and leads then to
turbulent diffusion of the field through mixing.

\begin{figure}[t]
\includegraphics[width=\columnwidth]{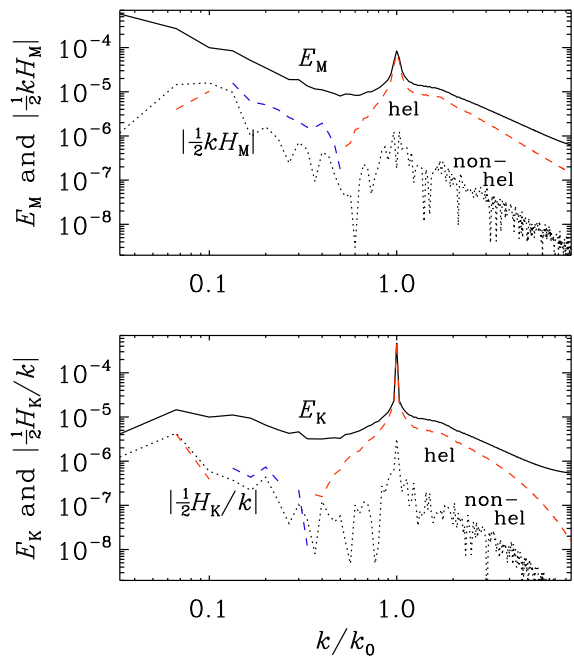}
\caption{
Comparison of normalized magnetic (upper panel) and kinetic (lower panel)
helicity spectra for cases with helical (dashed) and non-helical (dotted)
forcings.
In the case with helical forcing (dashed lines), red/blue segments
indicate positive/negative values of $H_{\rm M}$ and $H_{\rm K}$.
For comparison, we also show magnetic and kinetic energy spectra
in the upper and lower panels, respectively.
}\label{pspec2te}
\end{figure}

\subsection{Comparison with helical forcing}

As we have seen in \Fig{pspec2tb}, there is virtually no difference
between vortical forcings with and without helicity.
This is probably related to the fact that there was not enough time
for the helical forcing to affect magnetic and velocity fields at scales
larger than the injection scale.
This can be seen from \Fig{pspec2te}, where we compare cases with
helical and non-helical forcings for magnetic and kinetic helicity spectra,
$H_{\rm M}(k)$ and $H_{\rm K}(k)$, respectively.
These spectra are normalized such that
$\int H_{\rm M}(k)\,dk=\langle{\textbf A}\cdot{\textbf B}\rangle$ and
$\int H_{\rm K}(k)\,dk=\langle{\bm\omega}\cdot{\textbf u}\rangle$, where
${\bm\omega}={\bm\nabla}\times{\textbf u}$ is the vorticity.
These spectra are shown together with their respective magnetic and kinetic
energy spectra, $E_{\rm M}(k)$ and $E_{\rm K}(k)$, respectively.
They satisfy the realizability conditions, $E_{\rm M}\ge|\half kH_{\rm M}|$
and $E_{\rm K}\ge|\half H_{\rm K}/k|$, respectively \cite{Mof69}.
(Note that the definitions of $\int H_{\rm M}$ and $\int H_{\rm K}$
relative to those of $E_{\rm M}$ and $E_{\rm K}$ differ from each other
by a $k^2$ factor, which explains the slightly different from of their
respective realizability conditions.)

\FFig{pspec2te} shows that in the case with helical forcing, where the
kinetic helicity is positive, $\half k H_{\rm M}(k)$ is also positive
and close to $E_{\rm M}(k)$ for $k\approx k_0$, but about an order of
magnitude below it for smaller values of $k$.
This supports the suggestion that helicity effects are too small
to modify the energy spectra significantly.

\section{Conclusion}

In this paper we have studied the evolution of an inflation-generated
magnetic field \cite{Ratra:1992ab} coupled to the fluid during
cosmological phase transitions. Our formalism is very general and
applies to the electroweak and QCD phase transitions. The difference
between these (and other) phase transitions is encoded in the difference
in parameters such as the temperature and the number of relativistic
degrees of freedom, parameters which determine the characteristic length
scale of the system under consideration ($\lambda_0$). We consider
different types of forcing and show that at late times the kinetic energy
spectrum depends sensitively on the forcing used.

Our forcing scale is determined by the phase transition bubble size.
Within a few turnover times the kinetic energy spectrum starts to rise
on large scales, generating large-scale turbulent motions in the fluid.
Even a rapid phase transition generates turbulence, which will slowly
decay on large scales. Phase transition-generated MHD turbulence might
be relevant for cosmological magnetogenesis \cite{Widrow:2012ab}. Phase
transition turbulence can also generate a gravitational wave signal
that is potentially detectable \cite{e-LISA}.

In contrast to previous studies, the inflation-generated magnetic field
is not frozen into the cosmic plasma.
The forcing that we considered here is limited by the duration of the
phase transition. After the forcing source stops to act, both
magnetic and kinetic energies start to decay freely.
The configuration of the magnetic field at large scales (outside the
phase transition Hubble radius) is almost unchanged.
At intermediate scales corresponding to the phase transition
bubble size there is a slight suppression due to energy conversion
into kinetic energy.
The induced turbulent motions are causal so the spectral shape at large
scales is given by a white noise spectrum $E_K(k) \propto k^2$ \cite{hogan};
the vorticity energy density spectrum will be steeper ($k^4$) due to the
additional requirement of causality \cite{caprini2003}.
The presence of magnetic helicity does not significantly change
the forcing stage. On the other hand, the scaling laws in the decay stage
are strongly affected by the presence of magnetic helicity.
The duration of the decay stage is much longer than the forcing stage.
During this stage the correlation length of the velocity increases
with a corresponding decay of the total energy density.
The magnetic field on super-Hubble radius scales is decoupled
from the fluid which, in turn, stays almost unaffected.

The main results of our study are: (i) inflation-generated
magnetic fields are not significantly modified on large scales
by their coupling to the plasma during a cosmological phase transition;
(ii) the coupling of the magnetic field with the phase transition fluid
leads to deviations of the magnetic field spectrum from the initial
scale-invariant shape on intermediate scales; and, (iii) there is the
possibility of having large-scale correlated turbulent motions in the
early Universe which, eventually, could affect the development of
large-scale structure formation at late times, and in particular cluster
physics \cite{krav}.

 \acknowledgments

We appreciate useful discussions with C.\ Caprini, K.\ Jedamzik,
A.\ Kosowsky, A.\ Kravtsov, A.\ Neronov, and D.\ Semikoz.
Computing resources have been
provided by the Swedish National Allocations Committee at the
Center for Parallel Computers at the Royal Institute of Technology
in Stockholm and and by the Carnegie Mellon University supercomputer center.
We acknowledge partial support from DOE grant DEFG030-99EP41093,
Swiss National Science Foundation SCOPES grant 128040, NSF grants
AST-1109180 and AST-1109275, NASA Astrophysics Theory Program grant
NNXlOAC85G, European Research Council AstroDyn Research Project
227952, and Swedish Research Council grant 621-2007-4064.
T.K.\ acknowledges the ICTP associate membership program.
A.B.\ and A.T.\ acknowledge the hospitality of the McWilliams Center
for Cosmology where part of this work was performed.

\end{document}